\documentclass[runningheads]{llncs}
\usepackage[T1]{fontenc}
\usepackage{amsmath}
\usepackage{graphicx}
\usepackage{url}
\begin{document}
\title{A protocol to reduce worst-case latency in deflection-based on-chip networks\thanks{This preprint contains errors in the analysis and experimental evaluation that were identified after its initial submission. In particular, the proposed re-injection mechanism requires increased switch buffering that is not accounted for, and the dominance claim does not hold. A corrected analysis using an alternative re-injection strategy, along with updated experimental results, appears in the published version~\cite{published}. The published version supersedes this preprint.}}
\author{Leandro Soares Indrusiak\orcidID{0000-0002-9938-2920}} 
\authorrunning{L. Soares Indrusiak}
\institute{School of Computer Science, University of Leeds} 

\maketitle              % typeset the header of the contribution
\begin{abstract}
We present a novel protocol that reduces worst-case packet latency in deflection-based on-chip interconnect networks. It enforces the deflection of the header of a packet but not its payload, resulting in a reduction in overall network traffic and, more importantly, worst-case packet latency due to decreased pre-injection latency.     

\end{abstract}
\section{Introduction}

Deflection routing has been used in on-chip networks, most often in ring topologies, to manage network traffic efficiently while reducing hardware overheads~\cite{Ausavarungnirun2014}. Instead of relying on buffers to store packets when their desired network link is congested, deflection routing sends packets along alternative network paths, even if these paths are longer. By ensuring that packets keep moving continuously, it reduces the need for extensive buffering and complex flow control mechanisms. Previous work has shown that a deflection-enabled ring network can have an order of magnitude less energy dissipation and occupy up to 85\% less chip area than an equivalent mesh-based network~\cite{alazemi_routerless_2018}, making it an attractive architecture for embedded multiprocessor systems.

Performance guarantees are a key requirement in embedded multiprocessor platforms, and deflection can have a severe impact on network latency, which in turn affects end-to-end performance. To quantify that impact, worst-case latency models for on-chip networks with deflection routing have been produced for architectures that deflect individual flits of a packet~\cite{Ribot2022} or a complete packet~\cite{Indrusiak2023}. This paper addresses networks with full-packet deflection, a choice that we justify in more detail in Section \ref{background}.

The main contribution of the paper is presented in Section \ref{proposed}. It includes a novel protocol that can be used in full-packet deflection ring networks, and a respective worst-case analysis that quantifies the impact of the introduction of that protocol on the latency of network packets. Perhaps counter-intuitively, the proposed protocol does not directly impact the number of deflections suffered by a given packet, and therefore does not necessarily reduce the time between the packet's injection into a network and its ejection at destination. Instead, our protocol reduces the interference that a given packet causes to the injection of other packets, thus reducing their worst-case latency. 

The paper is closed with extensive experimental work, comparing worst-case performance guarantees provided by deflection networks such as~\cite{alazemi_routerless_2018} and ~\cite{liu_imr_2016} (using the analysis proposed in~\cite{Indrusiak2023}) against equivalent networks using the proposed protocol. Subsequent analysis has revealed errors in the proposed method; a corrected version with updated results has been published~\cite{published}.

\section{Background work}\label{background}

\subsection{Deflection routing in on-chip networks}

Deflection routing redirects traffic to alternative output ports of a network switch when their preferred path is unavailable, either due to congestion~\cite{Fallin2012} or network faults~\cite{Sleeba2018}. Besides the ability to avoid congestion or tolerate faults, deflection routing networks can be more resource efficient, requiring little or no buffering and simpler flow control logic. On the other hand, deflection routing can have a negative impact on packet latency and may introduce or increase the possibility of packet livelock or deadlock~\cite{Sleeba2018}.  

The resource efficiency of deflection routing networks makes them particularly well-suited for on-chip interconnects in embedded and real-time platforms, where constraints on area, power, and latency are critical. These advantages have been empirically demonstrated in previous work, which compared deflection-enabled ring networks against fully-buffered mesh-based networks: Liu et al. show significant improvements on latency and throughput~\cite{liu_imr_2016} as well as minor savings in area and energy dissipation, while Alazemi et al.~\cite{alazemi_routerless_2018} managed to reduce even further the area and energy footprint achieving results that are up to 80\% smaller and dissipate up to 10 times less energy than the baselines. Similarly, Wasly et al.~\cite{HopliteRT2017} exploit the buffer efficiency of deflection routing to propose an interconnect that can be economically implemented within an FPGA fabric.  

Resource efficiency is not the only requirement for embedded and real-time platforms, and the ability to guarantee the timeliness and performance of critical functions is at least just as important. Several works try to upper-bound the impact of deflection routing on the latency of time-critical communication in on-chip networks, and to do so they must focus on how deflection is performed. We can divide those approaches in two groups, addressing either flit-level or packet-level deflection:
\begin{itemize}

\item Flit-level deflection happens when the deflection decision is applied to each individual data unit within a packet (referred as a flit~\cite{Dally92}). This means that each flit of a packet may take a different path towards the destination, as congestion may occur only when some of the packet's flits are crossing a particular network switch~\cite{Ribot2020}\cite{HopliteRT2017}. Therefore, latency upper-bounds for such networks must consider the additional latency caused by flits following longer network paths, as well as the overheads of the mechanisms that handle or prevent out-of-order flit delivery~\cite{Ribot2020}\cite{Ribot2022}.

\item Packet-level deflection happens when the deflection decision can only be applied to the header flit of a packet, and in case the header is deflected its whole payload is deflected as well~\cite{liu_imr_2016}\cite{alazemi_routerless_2018}. This type of deflection maintains the order of flits within a packet, as the payload follows the header in a pipeline fashion, so the on-chip network does not incur any energy, area or performance penalty due to flit reordering mechanisms. Furthermore, it only requires routing information to be included in the header flit, as opposed to flit-level deflection which requires each and every flit to carry routing information (which in turn increases significantly the area and energy dissipated by link wires, switch multiplexers and registers). As a downside, packet-level deflection typically requires larger buffers that are able to accommodate a full packet. 

\end{itemize}

In this paper, we aim to benefit from the resource efficiency of packet-level deflection as reported in~\cite{alazemi_routerless_2018}, as that architecture does not require redundant routing information in every flit (smaller area and energy dissipation) or any flit reordering mechanism (smaller area, energy dissipation and latency). In the following subsections, we describe that architecture in more detail, and review the analytical models that can be used to upper-bound its latency.

\subsection{Packet-level deflection routing in routerless networks}\label{routerless_architecture}

Existing packet-level deflection networks rely on ring topologies, with one or more rings connecting the network switches. Ausavarungnirun et al.~\cite{Ausavarungnirun2014} used a hierarchical ring topology without any in-ring buffering. If packets needed to move across rings to reach their destination, they are deflected until inter-ring buffers become available. Liu et al.~\cite{liu_imr_2016} introduced IMR, a more minimal design where packets never cross rings. Instead, rings are statically defined to ensure full network connectivity. They propose a genetic algorithm to optimize ring configurations towards minimal latency and interconnect cost. Each switch includes a full-packet buffer per ring and a single ejection link shared across rings. Packets may be deflected and keep looping within their ring until the ejection link is free. Livelock is avoided using timestamps and an Oldest-First arbitration policy. Alazemi et al.\cite{alazemi_routerless_2018} improved IMR with RLrec, a simpler heuristic for ring placement, shared buffers across rings, and introduced the possibility of multiple ejection links per switch. Due to the simplified flow control and reduced buffering, such networks are often called routerless networks. 

\begin{figure}[h]
  \centering
  \includegraphics*[scale=1.0]{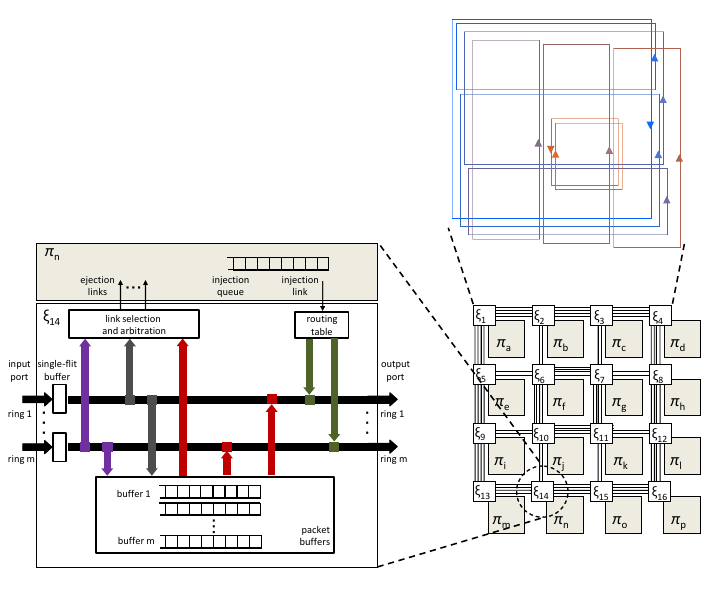}
  \caption{Detail of a routerless network switch architecture as proposed by~\cite{alazemi_routerless_2018} within a 4x4 network topology with 10 rings generated by the RLrec heuristic.}
  \label{fig:overview}
  %\vspace{-0.8cm}
\end{figure}

Figure~\ref{fig:overview} illustrates a 16-switch 10-ring routerless network created using RLrec. Each switch connects to a local processor core and to one or more rings, with injection/ejection links between switch and core, input and output ports between switches, and packet buffers in each switch. Flits coming into the input port of a switch are processed as they arrive, and can be either ejected (if the switch is their destination, and the ejection link is free), forwarded down the ring (if the switch is not their destination, or if it is but the ejection link is busy, so a deflection must take place), or buffered (if an injection is ongoing, or if the packet buffer is not empty). Flits can only be injected if there are no flits coming into the switch's input port and if the packet buffer is empty, ensuring ring traffic has priority. In other words: output ports prioritize flits from packet buffers, input port, or injection link, in that order, unless a packet is mid-injection. As a result, routerless switches require no flow control, as flits keep moving down their ring as downstream buffers are guaranteed to be available. 

\subsection{Upper-bounding packet latency in routerless networks}\label{background_analysis}

Latency upper-bounds for routerless networks were first proposed by Indrusiak and Burns in~\cite{Indrusiak2023}, with distinct formulations to accommodate network configurations where injection and ejection links are or aren't shared among rings. Before we can understand those formulations, we must review the notation established in~\cite{Indrusiak2023}, which we will reuse throughout the rest of this paper. 

\subsubsection{System model}

A routerless on-chip network such as those described in subsection \ref{routerless_architecture} consists of a set of processing cores $\Pi = \{\pi_a,\pi_b, \ldots, \pi_z\}$ and a set of rings $O = \{o_1,o_2, \ldots, o_m\}$. Each ring $o \in O$ is defined by an ordered set of $r^o$ switches $\Xi^o = \{\xi_1,\xi_2, \ldots, \xi_{r^o}\}$ and a set $\Lambda^o$ of $3r^o$ unidirectional links (these include a link between subsequent switches of the ring and two links per switch for communication with its local core, i.e. ejection and injection links). Each switch $\xi \in \Xi^o$ contains a buffer of size $B^o$, sufficient to store the largest packet that may be injected into the ring.

A switch may belong to multiple rings (e.g. $\xi_2 \in \Xi^{o^1}, \xi_2 \in \Xi^{o^3}, \xi_2 \in \Xi^{o^7}$, in which case in includes the links and buffers of each ring it participates in (see Figure \ref{fig:overview}). 

The network traffic is modelled as a set $\Gamma$ =$ \{\tau_1,
\tau_2, \ldots \tau_n\}$ of $n$
real-time traffic flows. Each flow $\tau_i$ generates a potentially unbounded sequence
of packets. A flow is characterised by the tuple
$\tau_i$ = ($T_i$, $D_i$, $L_i$, $J_i$, $\pi_i^s$,
$\pi_i^d$), where $\pi_i^s$ and $\pi_i^d$ are the source and destination cores, respectively. Flows are assumed to be periodic or sporadic, with $T_i$ denoting the minimum inter-arrival time (in cycles). The maximum packet size, in flits, is $L_i$, which also represents the time in cycles required to transmit a full packet across a link. Each flow has a relative deadline $D_i$ (in cycles), and a maximum release jitter $J_i$, also in cycles. The formulations in~\cite{Indrusiak2023} are based on the assumption that $D_i \le T_i$.

The formulations in~\cite{Indrusiak2023} also define the function $path^o(\pi_\alpha, \pi_\omega)$ to denote the ordered subset of $\Xi^o$ with the switches in the path between cores $\pi_\alpha$ and $\pi_\omega$;, the function $dpath^o(\pi_\alpha, \pi_\omega)$ to denotes the downstream path between those processing cores (i.e. the exact same ordered subset of $\Xi^o$ except for the first switch connected to $\pi_\alpha$), and that the absolute value of those functions denotes the number of switches in the respective path, e.g. $| path^o(\pi_\alpha, \pi_\omega) | = | dpath^o(\pi_\alpha, \pi_\omega) | + 1$. They also make use of the concept of maximum no-load latency of a packet flow $\tau_i$, represented by $C_i$ in most worst-case analysis models, which in this case can be obtained by adding the number of ring links between $\pi_i^s$ and $\pi_i^d$ (including injection and ejection ones) and the number of payload flits of the packet ($L_i-1$).

Finally, particular subsets of $\Gamma$ were defined to simplify the analyses presented in ~\cite{Indrusiak2023}. Let $\Gamma^o \subset \Gamma$ denote the set of flows assigned to ring $o \in O$. Each flow is mapped to a single ring, so the sets $\Gamma^o$ are mutually exclusive. For a given flow $\tau_i$, the subset $\Gamma_{in_i} \subset \Gamma$ denotes the set of flows using the same injection link as $\tau_i$. Similarly, the subset $\Gamma^o_{up_i} \subset \Gamma^o$ is defined as the set of flows that include $\tau_i$'s injection switch in their $path^o$ (i.e. they go through $\tau_i$'s injection switch even if they are never deflected).

\subsubsection{Analysis}

The latency upper-bound formulation in~\cite{Indrusiak2023} is based on classic response-time analysis, and identifies all types of timing interference suffered by the flow under analysis. The formulation exploits the non-preemptive nature of routerless networks, and separates the analysis in two distinct stages: before and after injection. 

The first stage of the analysis quantifies the worst-case interference before injection $I^{pre}$ suffered by a packet whilst waiting in the injection buffer before it can access its ring. That includes the time it is queuing behind packets from other flows ($I^{pre_{queue}}$), as well as the time spent at the head of the queue waiting for its ring to become free ($I^{pre_{idle}}$). 

Equation \ref{eq:ipre_idle} is the $I^{pre_{idle}}$ upper-bound presented in~\cite{Indrusiak2023}, and it calculates the largest busy period in the switch where the packet under analysis is injected. It accounts for interference from other packets that go through that switch even if they are not deflected (i.e. set $\Gamma_{up_i}^{o}$), as well as up to $maxloop$ deflections of those and all other packets using that ring (i.e. set $\Gamma^o$). It also accounts from indirect interference using the notion of interference jitter $J_j^k$, which can be safely calculated as $R_j - C_j$ as proposed by~\cite{Shi08}, or calculated iteratively as in~\cite{Indrusiak2023} for a tighter estimate.  

\begin{multline}\label{eq:ipre_idle}
I_i^{pre_{idle}} \ =  1 + \sum_{\tau_j \in \Gamma_{up_i}^{o} } { \left\lceil
{\frac{I_i^{pre_{idle}} + J_j  + J_j^k }{T_j} } \right\rceil \cdot L_j } + \\ \sum_{\tau_j \in \Gamma^o }   \sum\limits_{1}^{maxloop_j} { \left\lceil
{\frac{I_i^{pre_{idle}} + J_j  + J_j^k }{T_j} } \right\rceil \cdot L_j }
\end{multline}

Equation \ref{eq:ipre_queue} is the $I^{pre_{queue}}$ upper-bound from~\cite{Indrusiak2023}, and it simply adds up the injection time and queuing time of all packets that share the injection link used by the packet under analysis (and therefore could be queued upon its release), denoted by the set $\Gamma_{in_i}$. 

\begin{equation}\label{eq:ipre_queue}
I_i^{pre_{queue}} \ =  \sum_{\tau_j \in \Gamma_{in_i} } {(L_j + I_j^{pre_{idle}})}
\end{equation}

The worst-case interference before injection $I^{pre}$ of a particular flow is then obtained by the sum of its $I^{pre_{queue}}$ and $I^{pre_{idle}}$. 

The second stage of the analysis quantifies the worst-case interference after injection $I^{pos}$, as it crosses the ring toward its destination. It accounts for all interference caused by potential packet injections that occur in switches along its way, as well as $maxloop$ deflections caused by busy ejection links at its destination.

\begin{equation}\label{eq:ipos}
I_i^{pos} \ =  (|dpath^o(\pi_i^s,\pi_i^d)|  \cdot B^o ) + (maxloop_i \cdot r^o \cdot B^o) 
\end{equation}

Equation \ref{eq:ri} then uses the interferences from both stages to calculate the latency upper-bound $R_i$ for a packet flow $\tau_i$: 

\begin{equation}\label{eq:ri}
R_i \ =  C_i + r^o \cdot maxloop_i  + I_i^{pre} + I_i^{pos}
\end{equation}

We refer the reader to~\cite{Indrusiak2023} for more detailed explanation of the analysis, and for additional formulations covering architectures with ring-exclusive injection and ejection links (which we do not address in this paper, as the increased hardware overheads make them less likely to be adopted in practice).

\section{Proposed protocol}\label{proposed}
Unlike approaches such as \cite{Kunthara2022}, we do not attempt to change packet routes or reduce the number of deflections. Instead, we aim to reduce the overhead of a deflection by avoiding the unnecessary transfer of the packet payload around the ring. We make the following assumption, which is true for all routerless networks reviewed in subsection \ref{routerless_architecture}, and for most ring-based deflection networks: deflections occur within a single ring, so deflected packets will always pass through their injection switch before reaching again their destination switch (where they will either be ejected or deflected once more). 

\begin{figure}[h]
  \centering
  \includegraphics*[scale=1.0]{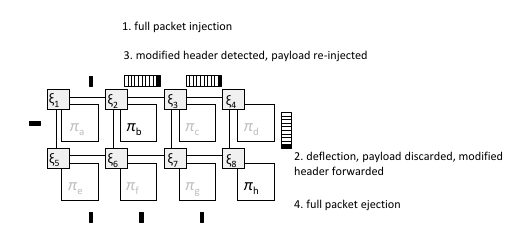}
  \caption{Proposed protocol applied to a 4x2 ring.}
  \label{fig:proposed}
\end{figure}

We therefore propose a protocol that discards the packet payload upon deflection, so only the packet header is deflected towards its injection switch, where the payload is then re-injected. Figure \ref{fig:proposed} depicts that approach applied to a scenario in which a packet is sent from $\pi_b$ to $\pi_h$ over a 4x2 ring. In step 1 the packet is fully injected by $\pi_b$ into the ring through the injection link of switch $\xi_2$. Then it is forwarded flit-by-flit across the ring via switches $\xi_3$ and $\xi_4$ until it reaches its destination switch $\xi_5$. Our approach is applied only in the case of deflections, so we consider a scenario where the packet cannot be ejected by $\xi_5$ at step 2 (e.g. its ejection link is shared with at least another ring that is not shown in Fig. \ref{fig:proposed}, and that ring is currently using the ejection link to deliver a different packet to $\pi_h$). At this point, the full packet would be deflected and go around the ring before attempting another ejection. Instead, our protocol deflects only the packet header and completely discards the packet payload. The header is then sent around the ring back to its injection switch $\xi_2$, which would then trigger $\pi_b$ to re-inject the payload to follow that header (step 3). The process from that point onward would be exactly the same as the original protocol until the packet reaches the destination switch. If the ejection link is available at that point, the packet is ejected as in the original protocol (step 4). If not, our protocol is reapplied, the payload discarded, the header is deflected back to the injection switch, and so on.

This is a very simple modification of the commonly used protocols in deflection networks, but it has significant consequences for the network's design, analysis and optimisation:

\begin{itemize}

\item The packet payload cannot be removed from the injection buffer immediately upon injection as it is commonly done in most networks, because it may have to be re-injected after each deflection. While this requires changes in the buffering logic (e.g. using SAFC or SAMQ buffering~\cite{Tamir88}), it does not require any additional buffering memory under the assumption that deadlines are constrained by the packet flow's period (i.e. $D_i \le T_i$, a requirement of the analysis in~\cite{Indrusiak2023}): a packet payload only needs to be deleted from the injection buffer upon the release of the subsequent packet of its flow, as it will occupy the same buffer space, but at that point the original payload would have already been delivered at the destination in a schedulable system, so no further re-injections would be needed.

\item The injection switch must be able to recognise the deflected header so that it can trigger a payload re-injection to follow that header. That process is subtly different from ejecting the header at the source switch and adding again the full packet to the injection queue, as that would go against the original network behaviour where deflected packets coming into a switch will prevent new injections. Instead, the header must be forwarded to the output port of the switch (or to the packet buffer if an injection is taking place), and the payload would then be injected flit by flit into the output port as soon as the header goes through it. Packet headers typically have enough storage space to store unique identifiers for their destination switch (so ejections happen at the right switch) as well as for their packet flow (so packets from different flows can be distinguished upon reception by the destination core). Those fields can be modified by destination switch upon deflection to indicate to the source switch that a re-injection must take place. The precise way to perform the field modification would depend on implementation-specific encoding and storage methods and is therefore outside the scope of the proposed protocol.  

\item A further modification could enforce that a modified header flit would be deflected even in case of a successful ejection, effectively enabling end-to-end acknowledgment which could support guaranteed delivery protocols (e.g. in radiation-tolerant interconnects). It could also be used to support the logic that manages the injection buffer, enabling earlier deletion of delivered payloads. 

\item By discarding the payload upon deflection, the proposed protocol reduces the amount of traffic, which can improve average latencies and the energy efficiency of the network. Those improvements are directly proportional to the size of the packets and the number of ring switches on the return path from the deflection point to the injection switch.  

\item The proposed protocol can be used in conjunction with techniques that aim to change routes or reduce deflections such as~\cite{Kunthara2022}. 

\end{itemize}

One metric is not directly affected by the proposed protocol: the actual packet latency. If considering a packet in isolation, the introduction of the protocol won't make a packet reach its destination any earlier or later. The discarding and re-injection of the packet payload does not change its own timing at all. What it does is to reduce the amount of interference it causes to other packets. By applying the protocol to the whole network we then expect worst-case latency improvements due to the overall reduction in timing interference. In the next section, we propose a modified version of the analysis reviewed in subsection \ref{background_analysis} that allows us to quantify those improvements.

\section{Worst-case analysis of the proposed protocol}\label{proposed_analysis}

Considering the separation of a packet latency in two stages as proposed in~\cite{Indrusiak2023}, we observe that the proposed protocol only affects the interference suffered by a packet before its injection, i.e. $I^{pre}$. Compared to the original network behaviour, the interference suffered by a given packet before its injection can be reduced significantly because instead of waiting for full deflected packets to flow out of its injection switch, it will instead wait only for their headers.

That change in behaviour can be quantified by changing the busy period formulation in Equation \ref{eq:ipre_idle}. The summation in the third term of the equation does not have to account for the deflections of all $\tau_j \in \Gamma^o$ flows. Instead, we need to account separately for deflected packets going from source to destination (i.e. full payload, either after injection or re-injection) and deflected packets going from destination to source (i.e. just a header, after the payload is discarded). We have already defined the set of flows crossing $\tau_i$'s injection switch whilst on their way from source to destination as $\Gamma^o_{up_i}$, so we can simply denote all the other flows crossing that switch after a deflection as the set $\Gamma^o_{def_i} = \Gamma^o - \Gamma^o_{in_i}$. We then rewrite Equation \ref{eq:ipre_idle} accordingly, to account for impact of the proposed protocol, as Equation \ref{eq:ipre_idle_new}. While a packet header is typically a single flit in most networks, we represent its length as $H$ for the sake of generality.

\begin{multline}\label{eq:ipre_idle_new}
I_i^{pre_{idle}} \ =  1 + \sum_{\tau_j \in \Gamma_{up_i}^{o} } { \left\lceil
{\frac{I_i^{pre_{idle}} + J_j  + J_j^k }{T_j} } \right\rceil \cdot L_j } + \\ \sum_{\tau_j \in \Gamma_{up_i}^{o} }   \sum\limits_{1}^{maxloop_j} { \left\lceil
{\frac{I_i^{pre_{idle}} + J_j  + J_j^k }{T_j} } \right\rceil \cdot L_j } +  \\ \sum_{\tau_j \in \Gamma_{def_i}^{o} }   \sum\limits_{1}^{maxloop_j} { \left\lceil
{\frac{I_i^{pre_{idle}} + J_j  + J_j^k }{T_j} } \right\rceil \cdot H }
\end{multline}

None of the other equations require any change, as the impact of the proposed protocol is fed into them through the updated values for $I^{pre_{idle}}$ produced by Equation \ref{eq:ipre_idle_new}.

We can prove by contradiction that the worst-case analysis of the proposed protocol dominates the analysis from~\cite{Indrusiak2023}. Considering that $\Gamma^o = \Gamma^o_{in_i} + \Gamma^o_{def_i}$, we can rewrite Equation \ref{eq:ipre_idle} and  separate its second summation into two separate summations over $\Gamma^o_{in_i}$ and $\Gamma^o_{def_i}$, so it resembles Equation \ref{eq:ipre_idle_new}. The only difference between the rewritten equation and Equation \ref{eq:ipre_idle_new} is the term $L_j$ instead of $H$ within the summation over $\Gamma^o_{def_i}$. Therefore, the original analysis would only dominate the proposed one if $L_j < H$, which is impossible (i.e. a full packet can never be smaller than its header).

\section{Evaluation}\label{eval}

We have established that the proposed protocol will always provide improvements on the worst-case latency of deflection-based networks. In this section, we aim to quantify the magnitude of that improvement over realistic ranges of network and packet sizes.

\subsection{Application-specific evaluation}

Firstly, we evaluate the magnitude of the reduction of the pre-injection latency for specific scenarios. The experimental setup consists of comparing the timing behaviour of a specific application when running over two distinct routerless networks-on-chip: one of them using the proposed protocol and another using the protocols used by Alazemi et al.~\cite{alazemi_routerless_2018} as a baseline. We use the 39-flow Autonomous Vehicle (AV) benchmark from~\cite{Indrusiak14}, configured for VGA resolution cameras, 8 bit color representation and 25 frames per second. To avoid biases, we generate 100 random mappings of the AV benchnmark to each of the networks, and compare the worst-case latencies of each of the flows for each mapping, using the analysis from~\cite{Indrusiak2023} for the baseline and the analysis described in this paper for the proposed protocol.

We configure both networks to use 32-bit flits, 1-flit headers, and the Oldest-First livelock prevention mechanism proposed by Liu et al.~\cite{liu_imr_2016} (which allows us calculate the value of $maxloop$ for each flow in each experiment as the number of flows competing for the same ejection link). 

Table \ref{tab:flowresults} below shows, for six different network topologies, the worst-case latency reduction achieved by proposed protocol as a percentage of the worst-case latency of the flow originally obtained by the baseline. The $max$ row shows the maximum improvement found for each network, and the $pm$ row shows the pooled mean improvement for each network. The pooled mean is calculated by averaging the improvement of all 39 flows of each mapping, and then averaging those values for all the 100 different mappings. 

\begin{table}[]
\caption{Reduction of worst-case latency achieved by the proposed protocol as a percentage of the baseline worst-case latency.}
\label{tab:flowresults}
\centering
\begin{tabular}{l|r|r|r|r|r|r}
\hline
\textbf{network}  & \textbf{4x4} & \textbf{5x5} & \textbf{6x6} & \textbf{7x7} & \textbf{8x8} & \textbf{9x9} \\ \hline
\hline
\textbf{max (\%)} & 93.07        & 89.45        & 89.26        & 89.33        & 83.36        & 80.66        \\ \hline
\textbf{pm (\%)}  & 6.60          & 3.33         & 3.20          & 2.64         & 2.16         & 0.92         \\ \hline
\end{tabular}
\end{table}

We can see that all topologies can benefit from the proposed protocol, and that it can provide worst-case latencies that are up to 93\% smaller than those achieved by the baseline. That happens for those flows where the pre-injection interference is most severe and therefore contributes significantly to their worst-case latency. As expected, the pooled mean shows a more modest improvement, as many flows may not encounter any interference, especially as the network scales: as the number of rings increases, the traffic generated by the AV benchmark is more evenly distributed, leaving less opportunity for improvement.

\subsection{Large-scale synthetic evaluation}

We now aim to evaluate the ability of the proposed protocol to improve schedulability across a large number of synthetic scenarios. We follow the flowset-based evaluation approach from~\cite{Indrusiak2023} and use \emph{schedulability ratio} as the main metric to compare the proposed protocol with the same baseline used in the previous subsection. Schedulability ratio is the percentage of cases, out of a set of benchmarks, that are deemed fully schedulable by a specific protocol and analysis: the worst-case latency of all its flows is less than their respective deadline. For a large and diverse set of synthetic benchmarks, one can argue that the protocol that achieves the highest schedulability ratio will likely produce a fully schedulable outcome in a practical network deployment. 

\begin{figure}[]
  \centering
  \includegraphics*[scale=0.43]{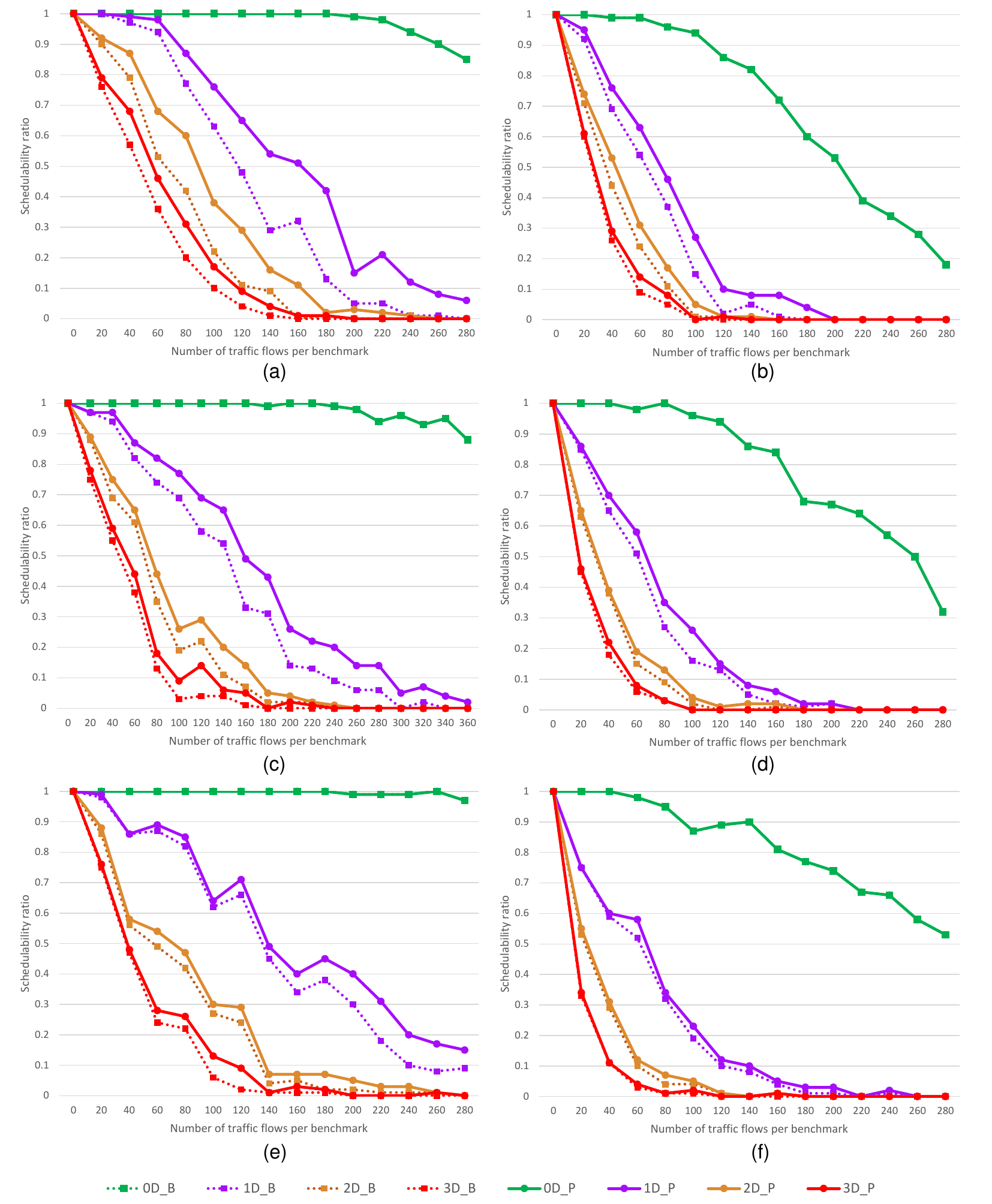}
  \caption{Comparative analysis based on schedulability ratio.}
  \label{fig:flowset}
\end{figure}

To compare the proposed protocol against the baseline, we synthetically generate a wide variety of benchmarks, each of them consisting of 100 randomly-generated flowsets. Each benchmark is characterised by the number of traffic flows per flowset. All experiments start with a benchmark containing 100 flowsets of 20 traffic flows each, and we compare the schedulability ratio of the proposed protocol against the baseline for that benchmark, i.e. how many of the 100 flowsets can be made fully schedulable (i.e. every single traffic flow in a flowset is schedulable). We then continue to generate 100-flowset benchmarks, but with more flows per flowset, up to a total of 280 flows, and perform the same comparison for each benchmark, aiming to evaluate the improvements achieved by the proposed protocol as the communication load increases.

The benchmarks are generated by uniformly sampling parameters from the following ranges: periods between 1 and 100 microseconds, release jitters between 0 and 50\% of the respective periods, and packet sizes between 16-48 flits or 32-96 flits. We randomly map those benchmarks upon networks of three different sizes (4x4, 5x5 and 6x6 cores), set to operate at a clock frequency of 1 GHz. For the sake of simplicity, we assign the same maximum number of deflections to all flows in every benchmark, from 0 to 3 deflections.  

Figure \ref{fig:flowset} plots the schedulability ratio results for three network sizes and two ranges of packet sizes: 
\begin{itemize}
\item 4x4 network and packets with 16-48 flits in Fig.\ref{fig:flowset}(a)
\item 4x4 network and packets with 32-96 flits in Fig.\ref{fig:flowset}(b)
\item 5x5 network and packets with 16-48 flits in Fig.\ref{fig:flowset}(c)
\item 5x5 network and packets with 32-96 flits in Fig.\ref{fig:flowset}(d)
\item 6x6 network and packets with 16-48 flits in Fig.\ref{fig:flowset}(e)
\item 6x6 network and packets with 32-96 flits in Fig.\ref{fig:flowset}(f)
\end{itemize}

Each plot shows schedulability ratio on the Y axis and the number of flows per flowset on the X axis. The plot tracks the schedulability ratio from smaller to larger benchmarks, so a higher line in the plot represents a superior protocol: it is able to maintain full schedulability in a larger percentage of the 100-flowset benchmarks even as their communication load (i.e. number of flows per flowset) increases. On each plot, we compare the proposed protocol (full lines) with the baseline (dashed lines) assuming that every flow in every benchmark is deflected 3, 2, 1 or 0 times:

\begin{itemize}
\item 0D\_B : baseline protocol, no deflection, analysis from~\cite{Indrusiak2023} 
\item 0D\_P : proposed protocol, no deflection, proposed analysis 
\item 1D\_B : baseline protocol, 1 deflection, analysis from~\cite{Indrusiak2023} 
\item 1D\_P : proposed protocol, 1 deflection, proposed analysis 
\item 2D\_B : baseline protocol, 2 deflections, analysis from~\cite{Indrusiak2023} 
\item 2D\_P : proposed protocol, 2 deflections, proposed analysis 
\item 3D\_B : baseline protocol, 3 deflections, analysis from~\cite{Indrusiak2023} 
\item 3D\_P : proposed protocol, 3 deflections, proposed analysis 
\end{itemize}

As expected, the highest schedulability ratio is found when flows suffer no deflection. In that case, the proposed protocol and the baseline behave exactly the same, so their respective analyses produce exactly the same worst-case bounds and their schedulability ratio is therefore the same (i.e. lines for 0D\_B and 0D\_P completely overlap). 

Then, we can see clear pairs of lines for the schedulability ratio under one, two and three deflections per flow in different colours. In every case, the proposed protocol performs better or at least as well as the baseline. The superiority of the proposed protocol is more pronounced for benchmarks with a single deflection, where it can achieve a better schedulability ratio by more than 20\%. In all cases, the proposed protocol achieve its best levels of improvement where the levels of load are not too low (where the baseline can also achieve full schedulability in most flowsets) or not too high (where most flowsets saturate the network).

The increase in packet size (plots b, d and f) also pushed the networks toward saturation, so the performance of the proposed protocol is better but with less than 10\% improvement over the baseline. 

The proposed protocol dominates the baseline as we increase size of the network, but the amount of improvement is more modest in larger networks. This is due to the nature of the RLrec algorithm used to generate the rings for each network, which provides a larger number of rings as the network scales up (e.g. 10 rings on a 4x4 network, 24 rings on a 6x6 network), many of them relatively small and handling local traffic (see Fig.\ref{fig:overview}), limiting potential interference and long deflections. We would expect better results in architectures with large rings such as IMR, but its use of long-running genetic algorithms to optimise ring layout prevented us from evaluating it in such a large-scale experiment.

\section{Conclusions}
We presented a novel protocol for deflection-based on-chip networks, aiming to reduce worst-case latency by eliminating the unnecessary deflection of packet payloads. We modified the state-of-the-art analysis to account for the timing behaviour of the proposed protocol, and showed that the proposed protocol dominates the protocol used in all state-of-the-art routerless deflection-based on-chip networks. It can achieve significant reduction of the worst-case pre-injection latency of specific flows (i.e. up to 90\% for those that are particularly affected by deflection traffic), and up to 6\% reduction on average across all flows.

The reduction of worst-case latencies is of particular interest in hard real-time systems, where the aim is to ensure systems are fully schedulable, i.e. all traffic flows can meet their deadlines even in the worst case. We have shown in experiments covering a large number of benchmarks that the proposed protocol dominates the baseline, and can provide fully-schedulable solutions for hundreds of cases deemed unschedulable by the baseline.

The proposed protocol achieves all the above-mentioned advantages without requiring any additional buffer memory overheads, and can be implemented using well-known buffering techniques. Besides reducing worst-case latencies as demonstrated in this paper, the protocol also reduces average-case latencies and energy dissipation,  however the quantification of those reductions is left for future work.

\paragraph{Note.} Subsequent analysis revealed that the method proposed above contains errors. The re-injection mechanism requires switch buffers larger than the baseline to accommodate multiple simultaneous re-injections at the same source switch, and even with correctly sized buffers, the resulting increase in post-injection interference ($I^{pos}$) negates the gains in pre-injection idle time ($I^{preidle}$). A corrected analysis using an alternative re-injection strategy that avoids the buffer increase, along with updated experimental results, appears in~\cite{published}. The published version supersedes this preprint.

\bibliographystyle{splncs04}
\bibliography{refs}
\end{document}